\documentclass[twocolumn,showpacs,twoside,10pt,aps,pra,superscriptaddress,longbibliography]{revtex4-2}
\usepackage{graphicx}
\usepackage{epsfig}
\usepackage{epsf}
\usepackage[utf8]{inputenc}
\usepackage{textcomp}
\usepackage{amsmath,amsfonts,amssymb,natbib}
\usepackage{url,textcase, bm}
\usepackage{amsthm}
\usepackage{multirow}
\usepackage{mathrsfs}
\usepackage{caption}
\usepackage{ragged2e}
\usepackage[colorlinks=true,citecolor=blue,urlcolor=blue, linkcolor=red]{hyperref}

\begin{document}
\title{
Scalable Fabrication-defined Multiplexed Optical Data Storage
}

\author{Junyu Guan}
\thanks{These authors contributed equally to this work.}
\affiliation{Laboratory of Spin Magnetic Resonance, School of Physical Sciences, Anhui Provincial Key Laboratory of Scientific Instrument Development and Application, University of Science and Technology of China, Hefei 230026, China}
\affiliation{Hefei National Research Center for Physical Sciences at the Microscale, University of Science and Technology of China, Hefei 230026, China.}

\author{Quanshen Shen}
\thanks{These authors contributed equally to this work.}
\affiliation{Laboratory of Spin Magnetic Resonance, School of Physical Sciences, Anhui Provincial Key Laboratory of Scientific Instrument Development and Application, University of Science and Technology of China, Hefei 230026, China}

\author{Bowen Tong}
\thanks{These authors contributed equally to this work.}
\affiliation{Laboratory of Spin Magnetic Resonance, School of Physical Sciences, Anhui Provincial Key Laboratory of Scientific Instrument Development and Application, University of Science and Technology of China, Hefei 230026, China}

\author{Hanzhi Wang}
\affiliation{Laboratory of Spin Magnetic Resonance, School of Physical Sciences, Anhui Provincial Key Laboratory of Scientific Instrument Development and Application, University of Science and Technology of China, Hefei 230026, China}

\author{Zeyu Gao}
\affiliation{Laboratory of Spin Magnetic Resonance, School of Physical Sciences, Anhui Provincial Key Laboratory of Scientific Instrument Development and Application, University of Science and Technology of China, Hefei 230026, China}

\author{Hanyu Zhang}
\affiliation{Laboratory of Spin Magnetic Resonance, School of Physical Sciences, Anhui Provincial Key Laboratory of Scientific Instrument Development and Application, University of Science and Technology of China, Hefei 230026, China}
\affiliation{Hefei National Laboratory, Hefei 230088, China}

\author{Jingyang Zhou}
\affiliation{Laboratory of Spin Magnetic Resonance, School of Physical Sciences, Anhui Provincial Key Laboratory of Scientific Instrument Development and Application, University of Science and Technology of China, Hefei 230026, China}

\author{Zihua Chai}
\affiliation{Laboratory of Spin Magnetic Resonance, School of Physical Sciences, Anhui Provincial Key Laboratory of Scientific Instrument Development and Application, University of Science and Technology of China, Hefei 230026, China}

\author{Dong Liu}
\affiliation{Laboratory of Spin Magnetic Resonance, School of Physical Sciences, Anhui Provincial Key Laboratory of Scientific Instrument Development and Application, University of Science and Technology of China, Hefei 230026, China}
\affiliation{School of Biomedical Engineering and Suzhou Institute for Advanced Research, University of Science and Technology of China, Suzhou, China}
\affiliation{Jiangsu Provincial Key Laboratory of Multimodal Digital Twin Technology, Suzhou, China}
\affiliation{Institute of Quantum Sensing of WuXi, Wuxi, China}

 \author{Ya Wang}
\affiliation{Laboratory of Spin Magnetic Resonance, School of Physical Sciences, Anhui Provincial Key Laboratory of Scientific Instrument Development and Application, University of Science and Technology of China, Hefei 230026, China}
\affiliation{Hefei National Research Center for Physical Sciences at the Microscale, University of Science and Technology of China, Hefei 230026, China.}
\affiliation{Hefei National Laboratory, Hefei 230088, China}

\author{Kangwei Xia}
\email{kangweixia@ustc.edu.cn}
\affiliation{Laboratory of Spin Magnetic Resonance, School of Physical Sciences, Anhui Provincial Key Laboratory of Scientific Instrument Development and Application, University of Science and Technology of China, Hefei 230026, China}
\affiliation{Hefei National Research Center for Physical Sciences at the Microscale, University of Science and Technology of China, Hefei 230026, China.}
\affiliation{Hefei National Laboratory, Hefei 230088, China}

\begin{abstract}
The accelerating growth of global data demands storage technologies combining high capacity, long lifetime and low energy consumption. Optical storage is attractive for archival applications, but conventional optical writing is constrained by diffraction-limited resolution and scalability, while thermal accumulation limits high-speed writing. Here we shift optical data writing from a serial optical process to a fabrication-defined process, in which lithographic masks govern where material modification occurs. Using patterned resist masks and ion implantation, we create fluorescent information voxels with nanoscale control over millimeter-scale areas. The platform supports multilevel and multidimensional encoding by independently encoding information in fluorescence intensity and wavelength. A machine-learning-assisted super-resolution reconstruction protocol is developed to recover information beyond conventional optical resolution for information readout. Overall, the framework reaches an areal storage capacity of 1.1 Tbit within the area of a standard 120-mm optical disc, approximately fivefold that of a conventional single-layer Blu-ray disc. This architecture establishes a micro/nanofabrication-compatible route to scalable optical memory and opens a pathway to wafer-scale archival storage.

\end{abstract}

\maketitle
% \linenumbers

\vspace{8pt}
\noindent{\fontfamily{phv}\selectfont 
\textbf{Introduction}}
\vspace{4pt}
\newline

The continued growth of digital information is intensifying the need for reliable long-term archival storage~\cite{315322}. Optical data storage is particularly attractive for such applications because of its low maintenance energy and long-term media stability~\cite{allison2026lasera, yuan2020ultrahigh}. However, conventional optical storage technologies, including CDs, DVDs and Blu-ray discs, remain constrained by diffraction-limited recording resolution and binary encoding, which limit both the minimum feature size and the amount of information stored within each addressable unit~\cite{sarid2007roadmap, gu2014optical}. These constraints fundamentally limit the storage density achievable with conventional optical media and have motivated the development of higher-density optical storage technologies~\cite{gu2016nanomaterials, lamon2021nanophotonicsenabled}.

Recent advances in optical data storage have substantially increased storage density beyond conventional optical-disc limits through sub-diffraction writing and multidimensional encoding~\cite{dhomkar2016longterm, monge2024reversible, huang2020reversible, sun2022threedimensional, zhang20253d}. Approaches based on write–deplete cycles~\cite{neupane2013tuning, dhomkar2016longterm, zhao20243d} and nonlinear optical effects ~\cite{huang2020reversible, lamon2021nanoscale, sun2022threedimensional, zhou2024terabitscalea, ye2025parallel} have enabled recording features below the conventional diffraction limit. Meanwhile, additional information dimensions, including wavelength, polarization and intensity, have been exploited to increase the information capacity within each spatial region~\cite{zijlstra2009fivedimensional, zhang2014seemingly, wang2022100layer, zhao20243d}. Despite these advances, many high-density schemes still form information voxels through localized optical modification of the storage medium~\cite{sun2022threedimensional, zhao20243d, zhou2024terabitscalea, allison2026lasera}. Serial optical addressing consequently limits writing scalability~\cite{ye2025parallel}, while thermal accumulation can constrain high-speed writing~\cite{allison2026lasera}.

Achieving both nanoscale writing resolution and scalable implementation therefore remains a central challenge for optical data storage. In conventional optical writing, spatial addressing and material modification are intrinsically coupled through the focused optical field. Separating these two functions offers an alternative route: micro- and nanofabrication can establish the spatial pattern independently, while ion implantation subsequently introduces optically active species over the patterned regions~\cite{herklotz2025modulating, ngandeungambou2024hot, telkhozhayeva2024roadmap}. This separation enables information voxels to be created through mask-defined material engineering, combining nanoscale spatial control with scalable implementation.

\begin{figure*}[t!]
\includegraphics[width=1.0\textwidth]{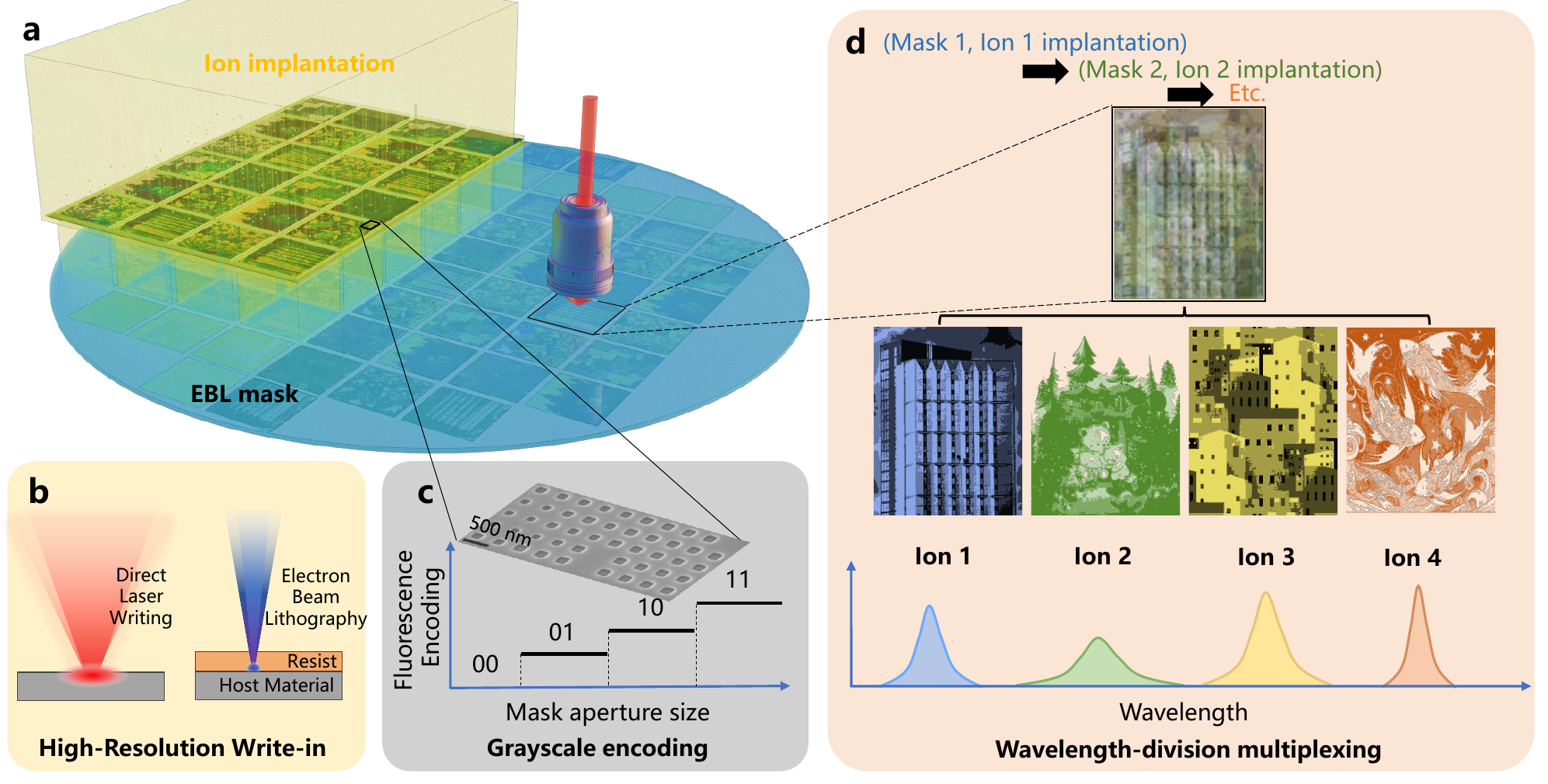}
\caption{
\textbf{Fabrication-defined optical data storage and multiplexing.}
\textbf{a.} Schematic of data writing using electron-beam lithography (EBL) and ion implantation.
\textbf{b.} Comparison of the spatial resolution of conventional laser direct writing and EBL.
\textbf{c.} Multilevel grayscale encoding through aperture-size-dependent fluorescence intensity (SEM image, scale bar: 500~nm).
\textbf{d.} Wavelength-division multiplexing using distinct implanted ion species and spectrally selective readout.
}\label{main-method}
\end{figure*}

Here we shift optical data writing from a serial optical process to a fabrication-defined process based on mask-defined material engineering. We implement this concept using electron-beam lithography (EBL) and ion implantation, combining nanoscale patterning beyond the diffraction-limited resolution of optical writing with chip-scale processing. The platform further supports both multilevel and multidimensional encoding by encoding information in fluorescence intensity and wavelength, respectively.  Machine-learning-assisted super-resolution reconstruction further enables the retrieval of densely encoded information beyond conventional optical resolution. Using this framework, we achieve an areal storage capacity of 1.1 Tbit within the area of a standard 120-mm optical disc. Collectively, these results establish a micro- and nanofabrication-compatible architecture for scalable, high-density optical memory and open a route towards wafer-scale optical memory fabrication.

\begin{figure*}[t!]
\includegraphics[width=1.0\textwidth]{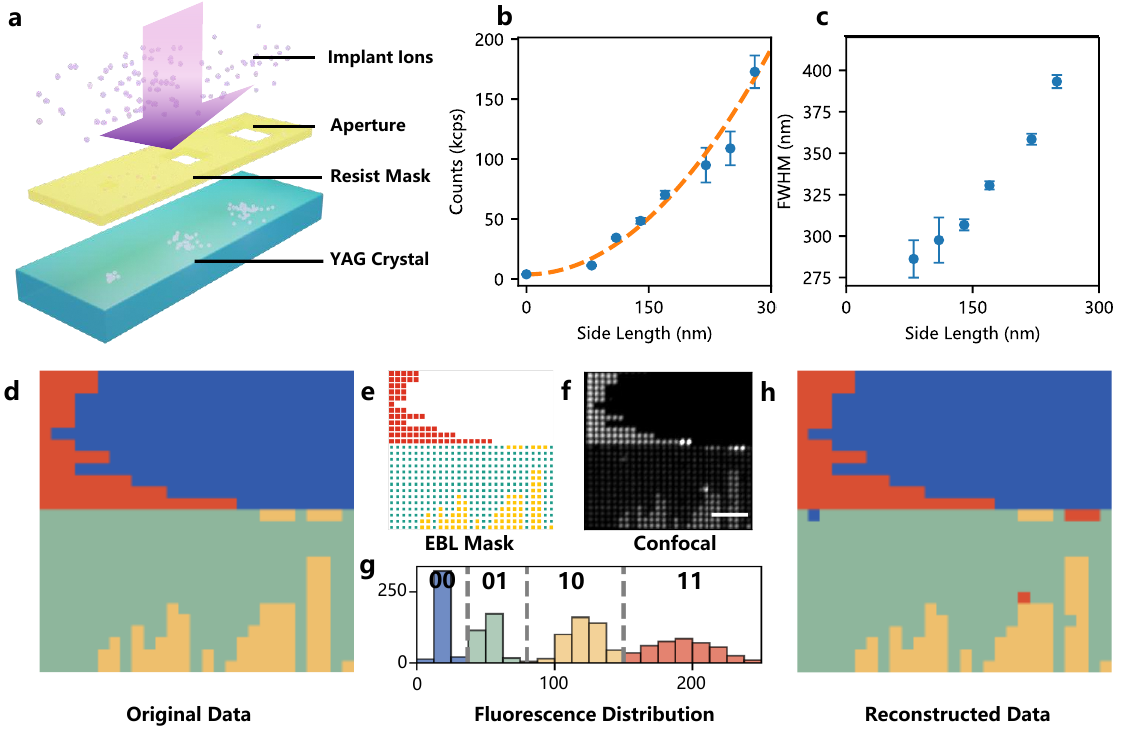}
\caption{
\textbf{Multilevel fluorescence encoding.}
\textbf{a.} Schematic of grayscale encoding using aperture-controlled ion implantation. The aperture size controls the number of ions implanted within each diffraction-limited voxel and thereby modulates the resulting fluorescence intensity.
\textbf{b.} Background-subtracted fluorescence counts as a function of mask aperture side length, together with a quadratic fit.
\textbf{c.} Fluorescence spot size as a function of mask aperture side length.
\textbf{d.} Target four-level grayscale data corresponding to 2-bit encoding.
\textbf{e.} EBL mask pattern for grayscale encoding with a voxel spacing of 500~nm. Grayscale level 0 corresponds to no implantation aperture, whereas levels 1, 2, and 3 are encoded using apertures with side lengths of 80~nm, 110~nm, and 140~nm, respectively.
\textbf{f.} Confocal fluorescence readout of the encoded pattern after Pr ion implantation and annealing (scale bar: 3~$\mu$m).
\textbf{g.} Fluorescence-intensity distributions extracted from individual voxel positions in \textbf{f}. Four distinct peaks resolve the four encoded grayscale levels.
\textbf{h.} Reconstructed grayscale data obtained from fluorescence-intensity decoding.
}\label{fig2}
\end{figure*}
 
\vspace{4pt}
\noindent{\fontfamily{phv}\selectfont 
\textbf{Results}}

Our fabrication-defined optical storage scheme decouples the spatial encoding of information from the physical formation of optically active information voxels. A resist layer is first lithographically patterned to encode the desired spatial layout. Global ion implantation is then performed through the resulting mask, followed by thermal activation of the implanted regions, as shown in Fig.~\ref{main-method}a. Compared with conventional laser direct writing, electron-beam lithography (EBL) provides nanoscale patterning beyond the diffraction-limited resolution of optical writing, as illustrated in Fig.~\ref{main-method}b.

Beyond reducing the spatial dimensions of information voxels, storage capacity can be further increased through both multilevel and multidimensional encoding. Within a diffraction-limited optical voxel, the mask aperture size controls the implanted ion number and consequently the fluorescence intensity, enabling multilevel grayscale encoding, as illustrated in Fig.~\ref{main-method}c. The encoding dimensionality can be further expanded by introducing wavelength as an independent degree of freedom. Multiple implantation steps with spectrally distinct species provide independently addressable wavelength channels within the same spatial region, enabling wavelength-division multiplexing, as illustrated in Fig.~\ref{main-method}d. Detailed fabrication procedures, mask design, and implantation simulations are provided in the Methods and Supplementary Notes 1 and 2.

For the experimental realization, we use yttrium aluminium garnet (YAG, Y$_3$Al$_5$O$_{12}$) as the storage host and rare-earth ions as the implanted optical emitters. YAG possesses high mechanical hardness, with a Mohs hardness of approximately 8.5. Its wide electronic bandgap underpins a broad transparency window extending from the ultraviolet to the near-infrared. Its chemical durability, together with its thermal and mechanical stability, provides a robust material platform for long-term information preservation. Unlike approaches in which information is encoded in exposed surface nanostructures, the optically active information in our scheme is embedded within a robust crystalline host. More importantly, the large family of rare-earth ions offers a broad range of spectrally distinct optical transitions, providing a natural basis for extending the number of independently addressable wavelength channels. Many trivalent rare-earth ions exhibit shielded $4f$-$4f$ transitions with high spectral stability and narrow optical linewidths, which are advantageous for reliable wavelength-selective encoding and readout.

In the present implementation, Ce$^{3+}$ and Pr$^{3+}$ are selected as two representative spectral channels. Both species can be efficiently addressed in the blue--violet spectral range and exhibit distinct excitation and emission characteristics, enabling selective optical readout within the same YAG host. Following implantation and thermal activation, rare-earth ions can occupy Y$^{3+}$ lattice sites with reported formation yields of 50--90\% ~\cite{kornher2016production, kolesov2018superresolution}, supporting efficient creation of optically active information voxels. Their relatively strong optical transitions further enable fluorescence detection at low ion numbers, down to the single-ion level~\cite{kolesov2012optical,kolesov2013mapping,xia2015alloptical}. The measured saturation count rates reach approximately 22~kcounts/s for single Ce$^{3+}$ ions and 1~kcount/s for single Pr$^{3+}$ ions in our platform. Further optical characterization and readout details are provided below and in Supplementary Note~3.

\begin{figure*}[t]
    \includegraphics[width=1.0\textwidth]{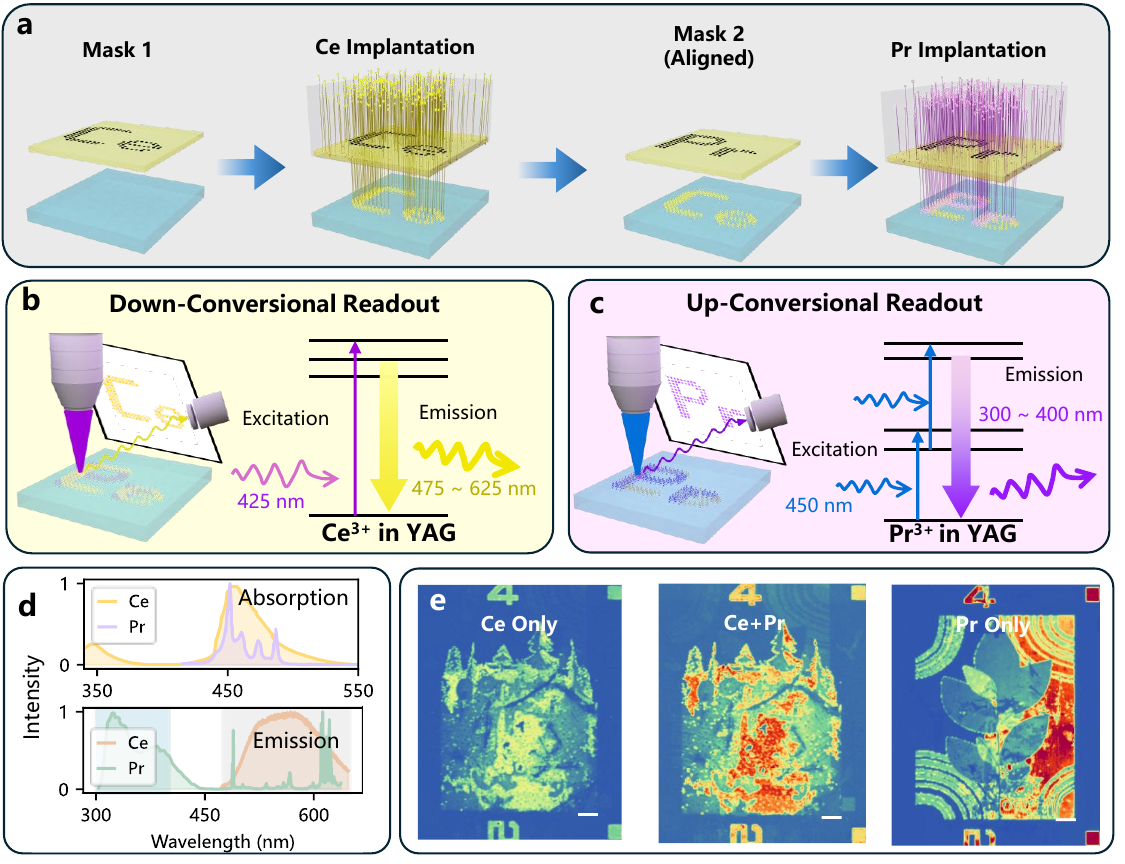}
\caption{\textbf{Wavelength-multiplexed fluorescence encoding.}
    \textbf{a.} Schematic illustration of the fabrication process for the wavelength-multiplexed sample. Ce ions are first implanted through Mask 1. After removal of Mask 1, an alignment procedure is performed to define Mask 2, through which Pr ions are subsequently implanted, enabling spatially registered encoding of two ion species within the same YAG crystal.
    \textbf{b.} Ce$^{3+}$ down-conversion fluorescence readout under 425~nm excitation, with fluorescence collected over 475--625~nm.
    \textbf{c.} Pr$^{3+}$ up-conversion fluorescence readout under 450~nm excitation, with fluorescence collected over 300--400~nm.
    \textbf{d.} Absorption and emission spectra of Ce- and Pr-doped YAG~\cite{jacobs1978measurement, gayen1992twophoton}, with the shaded regions indicating the spectral windows used for selective Ce$^{3+}$ and Pr$^{3+}$ readout.
    \textbf{e.} Experimental fluorescence images demonstrating species-selective readout of the multiplexed patterns. Left: selective readout of the Ce-implanted pattern under 425~nm excitation. Middle: visible-channel readout under 450 nm excitation, showing spectral crosstalk between the Ce- and Pr-implanted regions. Right: selective readout of the Pr-implanted pattern through up-conversion fluorescence under 450~nm excitation, with negligible crosstalk from Ce (scale bar: 4~$\mu$m).}
    \label{main-wl}
\end{figure*}

\begin{figure*}[t]
    \includegraphics[width=1.0\textwidth]{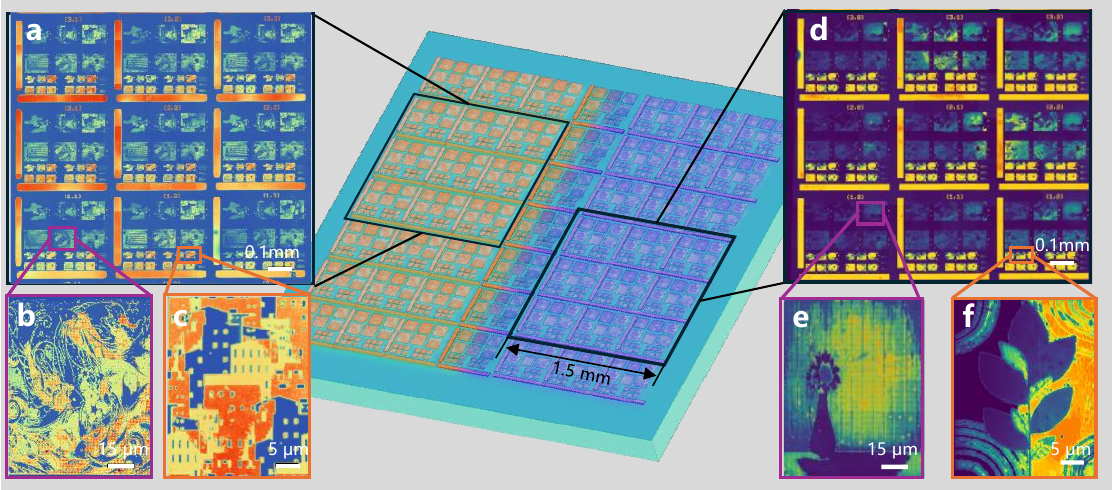}
    \caption{\textbf{Chip-scale fabrication-defined optical data storage.}
    \textbf{a.} Overview fluorescence image of the Ce-implanted  storage array (scale bar: 0.1~mm)
    \textbf{b.} Magnified fluorescence image of the Ce-implanted array with a voxel spacing of 500~nm (scale bar: 15~$\mu$m).
    \textbf{c.}  Magnified fluorescence image of the Ce-implanted array with a voxel spacing of 200~nm (scale bar: 5~$\mu$m).
    \textbf{d.} Overview fluorescence image of the Pr-implanted  storage array (scale bar: 0.1 mm)
    \textbf{e.}  Magnified fluorescence image of the Pr-implanted array with a voxel spacing of 500~nm (scale bar: 15~$\mu$m).
    \textbf{f.}  Magnified fluorescence image of the Pr-implanted array with a voxel spacing of 200~nm (scale bar: 5~$\mu$m).
}\label{large}
\end{figure*}

The single-ion detection capability demonstrated above provides the sensitivity required to resolve variations in implanted ion number through fluorescence readout, enabling multilevel information encoding. The encoding principle is illustrated in Fig.~\ref{fig2}a, where the aperture size of the patterned mask controls the mean number of ions implanted within each information voxel and thereby its fluorescence intensity.

To quantify this relationship, square apertures with side lengths ranging from 80 to 280~nm were patterned by EBL and used for Pr implantation into YAG. The background-subtracted fluorescence counts increase approximately quadratically with the aperture side length, consistent with the expected scaling of the implanted ion number with aperture area, as shown in Fig.~\ref{fig2}b. The full width at half maximum (FWHM) of the fluorescence spots also increases systematically with aperture size, as illustrated in Fig.~\ref{fig2}c. These measurements establish a reproducible mapping from lithographic mask geometry to fluorescence response, providing the calibration required for multilevel encoding.

The calibrated aperture-fluorescence relation is then used to implement four-level grayscale encoding, corresponding to 2-bit information per voxel. The target dataset contains four grayscale levels, labelled 0--3 and corresponding to the binary states 00, 01, 10, and 11, as shown in Fig.~\ref{fig2}d. The corresponding EBL mask was designed with a voxel spacing of 500~nm, where level 0 contains no implantation aperture and levels 1--3 are encoded using aperture side lengths of 80, 110, and 140~nm, respectively, as shown in Fig.~\ref{fig2}e. After Pr implantation and thermal activation, the encoded pattern was retrieved using confocal fluorescence microscopy, as shown in Fig.~\ref{fig2}f.

The fluorescence-intensity distributions extracted from individual voxel positions exhibit four distinguishable peaks corresponding to the four encoded levels, as shown in Fig.~\ref{fig2}g. Using intensity thresholds derived from these distributions, the original grayscale information is reconstructed in Fig.~\ref{fig2}h. The remaining decoding errors mainly arise from partial overlap between adjacent fluorescence-intensity distributions. Increasing the aperture-size separation between adjacent grayscale levels could further separate their fluorescence distributions and reduce decoding errors. Beyond the four levels demonstrated here, additional fluorescence levels could further increase the information encoded per voxel, provided that adjacent intensity distributions remain sufficiently separated for reliable decoding.

Beyond multilevel fluorescence encoding, the encoding dimensionality can be further expanded by introducing wavelength as an independent information channel. To demonstrate wavelength-division multiplexing, two independently patterned information channels were implemented through sequential Ce and Pr ion implantation through spatially aligned masks, as illustrated in Fig.~\ref{main-wl}a. The two ion species can be selectively addressed owing to their distinct absorption and emission spectra. In particular, Ce$^{3+}$ exhibits a broad $4f\rightarrow5d$ absorption band, whereas Pr$^{3+}$ is characterized by relatively narrow intra-$4f$ transitions ~\cite{jacobs1978measurement,gayen1992twophoton}, as shown in Fig.~\ref{main-wl}d.

Independent retrieval of the two channels is achieved by selecting appropriate excitation wavelengths and fluorescence detection windows. For Ce$^{3+}$ readout, 425~nm excitation is used, and the resulting down-conversion fluorescence is collected over 475--625~nm, as illustrated in Fig.~\ref{main-wl}b. For Pr$^{3+}$ readout, 450~nm excitation drives the $\mathrm{{}^{3}H_{4}\rightarrow{}^{3}P_{2}}$ transition, followed by further excitation to the $4f5d$ states. The resulting up-conversion fluorescence is detected over 300--400~nm, as shown in Fig.~\ref{main-wl}c. These complementary excitation and detection pathways provide spectrally distinct routes for retrieving the two encoded information channels.

\begin{figure*}
\includegraphics[width=1.0\textwidth]{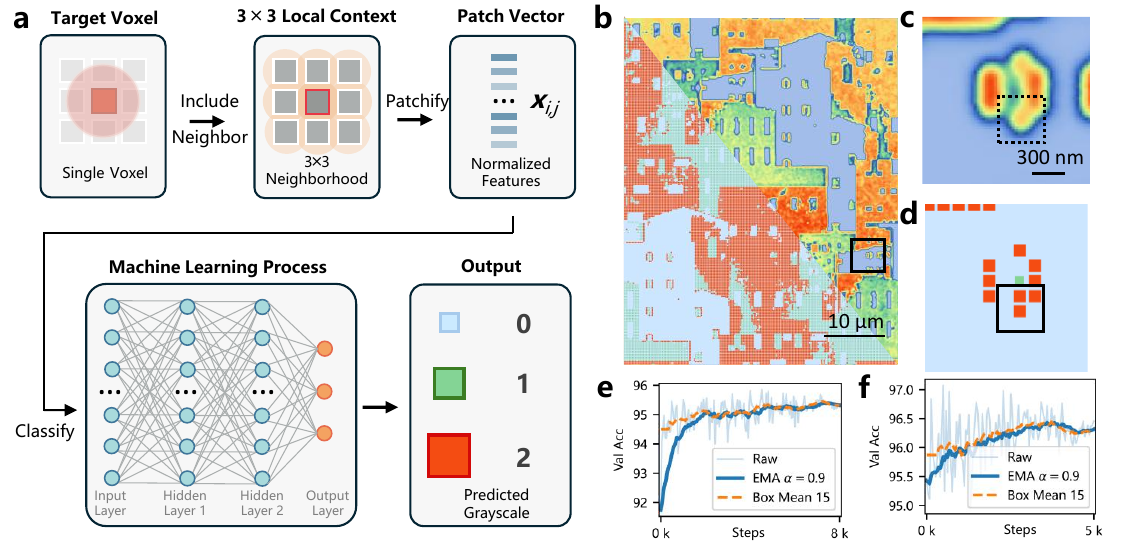}
\caption{\textbf{Machine-learning-assisted super-resolution reconstruction.}
\textbf{a.}  Workflow of the neighborhood-aware neural-network decoder for voxel-level classification. 
\textbf{b.} Fluorescence readout image of a high-density information array with a 200~nm voxel spacing. The dashed box indicates the region selected for local analysis (scale bar: 10~$\mu$m).
\textbf{c.} Magnified raw fluorescence image of the selected region, showing strong overlap between adjacent voxels (scale bar: 300 nm).
\textbf{d.} Ground-truth encoded voxel pattern corresponding to the region in \textbf{c}.
\textbf{e, f.} Validation accuracy curves for grayscale \textbf{e} and binary \textbf{f} classification. Raw validation accuracy, exponentially moving average smoothing ($\alpha$=0.9), and box moving average smoothing with a window size of 15 are shown to evaluate training stability and classification performance.
}\label{algorithm}
\end{figure*}

The wavelength selectivity is experimentally verified by laser-scanning fluorescence imaging, as shown in Fig.~\ref{main-wl}e. Under 425~nm excitation with detection over 475--625~nm, the Ce-implanted pattern is preferentially retrieved. At equal excitation powers, the Ce-implanted region produces a fluorescence signal of approximately 220~kcounts/s, whereas the Pr-implanted region yields approximately 36~kcounts/s. In comparison, 450~nm excitation with the same visible detection window produces substantial fluorescence from both implanted regions, with signals of approximately 420~kcounts/s and 270~kcounts/s for the Ce- and Pr-implanted regions, respectively. This comparison shows that excitation wavelength alone does not provide sufficient discrimination for selective Pr$^{3+}$ retrieval in the visible channel. Selective Pr$^{3+}$ readout is instead achieved by collecting its up-conversion fluorescence in the 300--400~nm ultraviolet window under 450~nm excitation. No detectable fluorescence from the Ce-implanted region is observed in this channel, enabling independent retrieval of the Pr-encoded pattern. These results show that independently addressable wavelength channels can be constructed through the combined selection of implanted species, excitation wavelength and detection window. The present two-species implementation could in principle be extended to additional rare-earth ions with spectrally distinct excitation and emission pathways, providing a route towards a larger number of independently addressable wavelength channels.

The scalability of the fabrication-defined architecture was next examined at the millimeter scale, as schematically illustrated in Fig.~\ref{large}. A single YAG sample was patterned with two spatially registered $3.5\times3.5$~mm$^2$ storage arrays within the same area, which were subsequently used for Ce and Pr ion implantation, respectively. Each array contained regions with different voxel spacings and encoding patterns, allowing multilevel, wavelength-multiplexed and high-density readout to be evaluated within the same fabricated device. Despite the millimeter-scale patterned area, nanoscale lithographic features were preserved, as confirmed by representative SEM images of the developed resist mask in Supplementary Note~1.

Fluorescence imaging confirms successful optical activation and readout across the millimeter-scale patterned regions for both Ce- and Pr-implanted arrays, as shown in Fig.~\ref{large}a and Fig.~\ref{large}d. Within these large-area arrays, regions were fabricated with voxel spacings of 500~nm and 200~nm for both implanted species, as illustrated in Fig.~\ref{large}b,c and Fig.~\ref{large}e,f. These results demonstrate that millimeter-scale processing can be combined with nanoscale lithographic control within a common fabrication workflow.

The reliability of information retrieval was then quantified for the 500~nm-spacing arrays. Using the readout and decoding procedures detailed in Supplementary Note~4, binary positional decoding achieved fidelities of 99.89\% and 99.64\% for the Pr- and Ce-implanted samples, respectively. With four-level grayscale encoding, the information capacity per voxel is doubled, while decoding fidelities of 94.26\% and 97.69\% are retained for the Pr- and Ce-implanted samples, respectively. The reduced grayscale fidelity is consistent with the partial overlap between adjacent fluorescence-intensity distributions discussed above, whereas residual positional errors are mainly associated with surface contamination or localized mask damage during processing.

Although EBL provides patterning resolution on the scale of tens of nanometers, conventional fluorescence readout remains limited by optical diffraction. In the 200-nm array, adjacent information voxels can no longer be reliably distinguished by conventional fluorescence imaging. This limitation is illustrated in the top-right region of Fig.~\ref{algorithm}b. The optical readout resolution therefore becomes the limiting factor for fully exploiting the spatial resolution afforded by the fabrication process. The use of fluorescent emitters as information voxels provides a natural interface to super-resolution fluorescence imaging. However, conventional super-resolution techniques generally require more complex and time-consuming image acquisition, which is unfavorable for large-area data readout. We therefore employ machine-learning-assisted reconstruction to recover densely encoded information directly from diffraction-limited fluorescence images.

As shown in Fig.~\ref{algorithm}b, the fluorescence image acquired at this spacing exhibits strong spatial overlap between adjacent voxels. The locally magnified image in Fig.~\ref{algorithm}c further shows that the measured intensity around a target voxel contains substantial contributions from neighboring sites. The corresponding encoded voxel pattern in Fig.~\ref{algorithm}d serves as the ground-truth reference for model training and decoding evaluation.

Instead of classifying each voxel based only on its center intensity, the neural-network decoder is designed to incorporate local spatial context, as illustrated in Fig.~\ref{algorithm}a. For each target voxel, the surrounding $3\times3$ local neighborhood is considered to capture the fluorescence coupling induced by adjacent voxels. The local fluorescence region is patchified and normalized into an input feature vector, which is then fed into a fully connected neural-network classifier to predict the grayscale state of the central voxel. By learning the correspondence between local fluorescence configurations and the underlying encoded voxel patterns, the model enables reliable voxel-level decoding under strong inter-voxel overlap, effectively achieving super-resolution readout beyond the diffraction limit.

Despite the strong spatial overlap in this sub-diffraction regime, the neighborhood-aware decoder recovers the encoded information from fluorescence patterns in which individual voxels cannot be directly resolved. The validation accuracy reaches 96.57\% for four-level grayscale decoding and 97.58\% for binary decoding, as shown in Fig.~\ref{algorithm}e,f. These results demonstrate that computational reconstruction can recover densely encoded information that is otherwise inaccessible through direct intensity-based readout.

The successful decoding of information at this spacing allows the spatial density enabled by the fabrication process to be utilized despite the diffraction-limited optical readout. At this spacing, the density is 2.5~Gbit/cm$^2$. Within the same fabricated storage platform, 200~nm voxel spacing is combined with four-level grayscale decoding and two independently addressable wavelength channels. These encoding methods correspond to an areal storage density of 10~Gbit/cm$^2$, equivalent to approximately 1.1~Tbit of storage within the footprint of a standard 120-mm optical disc and about fivefold the capacity of a conventional single-layer Blu-ray disc.

\vspace{8pt}
\noindent{\fontfamily{phv}\selectfont 
\textbf{Discussion}}
\vspace{4pt}
\newline
This work establishes a fabrication-defined optical storage architecture that shifts optical data writing from localized optical modification to programmable material engineering. Lithographic patterning and ion implantation provide control over both the spatial arrangement and optical response of information voxels, allowing spatial density, fluorescence intensity and spectral identity to be jointly engineered during fabrication. With machine-learning-assisted super-resolution reconstruction, the storage capacity corresponds to approximately 1.1~Tbit within the footprint of a standard 120-mm optical disc.

The demonstrated multilevel and multidimensional encoding can each be extended beyond the present implementation. Finer control of implantation aperture size could provide a larger number of resolvable fluorescence levels and thereby increase the number of bits encoded within each spatial voxel. In parallel, incorporating additional optically distinct ion species could introduce further wavelength channels within the same spatial region, providing another route to increase storage density. More broadly, the strategy could be extended beyond YAG to other solid-state platforms in which optically active centers can be engineered through fabrication. Further advances in large-area fabrication and parallel optical readout could ultimately extend this approach towards wafer-scale optical memory.

\vspace{8pt}
\noindent{\fontfamily{phv}\selectfont 
\textbf{Methods}}
\vspace{6pt}
\noindent{\fontfamily{phv}\selectfont 
\newline
\textbf{Sample Fabrication Process}}
\vspace{4pt}
\newline
The fabrication protocol for the large-scale ODS device is illustrated in the Supplementary Material Figure~S1. The procedure is as follows:
1.  A substrate is prepared by spin-coating an e-beam resist onto the host material.
2. Soft-bake the spin-coated e-beam resist. 
3. Expose the resist via EBL with the first information layer (Layer 1) to define a pattern corresponding to  Layer 1 of encoded information.
4. Develop the exposed e-beam resist. 
5. Implant the first species of ions through Layer 1 into the substrate. The patterned resist blocks ions, while the unmasked areas allow ions to penetrate the crystal, forming the first set of storage voxels.
6. Remove the residual resist, and a fresh layer of e-beam resist is spin-coated for a second EBL step, which encodes a new data pattern (Layer 2).
7. Implant the second species of ions through Layer 2 to create a new, spectrally distinct set of voxels. Repeat the above steps if more species of ions are required. 
8.  Remove the residual e-beam resist and anneal the host crystal to activate the fluorescence centers.
9.  The stored data is retrieved via wavelength-division multiplexing, reading the spectrally distinct voxel sets independently. 

\noindent{\fontfamily{phv}\selectfont 
\newline
\textbf{Sample Fabrication process}}
\vspace{4pt}
\newline
Undoped $\langle111\rangle$-oriented YAG single-crystal substrates (Ganzhou Qiandong Laser Technology Co., Ltd.) were used throughout this work. For single-ion imaging experiments, ultra-high-purity YAG crystals (Scientific Materials) were employed, with an impurity concentration below 36~ppb relative to the yttrium ion concentration in the YAG lattice.

Electron-beam lithography was performed using an EBL-ELS-F100 system (Elionix). The PMMA A4 (MicroChem)resist layer was spin-coated at 4000~rpm with an acceleration of 800~r/s$^{2}$ for 40~s, followed by soft baking at 180~$^\circ$C for 180~s. To mitigate charging effects during exposure, a conductive polymer layer (AR-PC~5090.02, ALLRESIST) was spin-coated under identical conditions and soft-baked at 120~$^\circ$C for 120~s. The mask patterns used for information writing were characterized by scanning electron microscopy (SEM) using a Crossbeam~550 system.

Ion implantation was carried out using a Metal Vapor Vacuum Arc (MEVVA) ion source, with the implantation angle fixed at 30$^\circ$. Ce ions were implanted with an energy of 40~keV and a dose of $3 \times 10^{14}~\mathrm{ions/cm^{2}}$, while Pr ions were implanted at 60~keV with a dose of $1.5 \times 10^{15}~\mathrm{ions/cm^{2}}$. Following implantation, the electron-beam resist was removed by acetone rinsing.

After implantation, the samples were annealed in air in a tube furnace (GSL-1400X-60, Hefei Kejing Materials Technology Co., Ltd.). The temperature was ramped at 5~$^\circ$C/min to 1200~$^\circ$C, held at 1200~$^\circ$C for 8~h, and then cooled to room temperature at a rate of 5~$^\circ$C/min.

\noindent{\fontfamily{phv}\selectfont 
\newline
\textbf{Machine Learning-Based Super-resolution Method}}
\vspace{4pt}
\newline
Super-resolution decoding was performed using a neural-network classifier that infers the storage state of each target voxel from its local fluorescence pattern. Unlike direct intensity-threshold decoding, which becomes unreliable when adjacent fluorescence signals strongly overlap, the classifier uses the spatial intensity distribution within the local neighborhood as its input.

Each experimentally acquired image was independently processed by histogram-based intensity remapping to compensate for fluorescence variations between acquisitions, followed by normalization to the range $[-1,1]$. The normalized images were partitioned into $15\times15$-pixel patches centered on each target storage unit, corresponding to the theoretical spatial extent of a $3\times3$ storage-unit neighborhood. These patches were used as direct inputs to the neural-network classifier.

The results reported in the main text were obtained using a three-layer multilayer perceptron trained with cross-entropy loss and the Adam optimizer. Model performance was evaluated using figure-level cross-validation, in which each experimentally acquired image was held out once for validation. Each fold was trained for 10 epochs on an NVIDIA RTX 4070 GPU and required approximately 1.5~min. A comparison of different model architectures is provided in Supplementary Note~5.

\vspace{8pt}
\noindent  {\fontfamily{phv}\selectfont 
\normalsize \textbf{Acknowledgements} 
}
\newline
\noindent This work is supported by the Strategic Priority Research Program of the Chinese Academy of Sciences (XDA0590204, XDB1740301), the National Natural Science Foundation of China (12274400, 12504594, T2325023, 92265204, 123B1019), Quantum Science and Technology-National Science and Technology Major Project (2021ZD0302200), and the Fundamental Research Funds for the Central Universities (WK2030000076). The nanofabrication was carried out at the USTC center for Micro and nanoscale Research and fabrication.

\vspace{8pt} 
\noindent  {\fontfamily{phv}\selectfont 
\normalsize \textbf{Author contributions} 
}
\newline
\noindent K.X., J.G., and Q.S. designed the experiment. Q.S., J.G., and H.Z. prepared the sample. J.G., Q.S., H.W., Z.C., Z.G. and K.X. performed the data-reading experiments and data analysis. B.T. and D.L. performed the algorithmic analysis. J.G., K.X., D.L., B.T., Q.S., Y.W., Z.C., J.Z., and Z.G. wrote the paper.  K.X. supervised the project and conceived the idea. All authors discussed the results and commented on the paper.

\vspace{8pt}
\noindent  {\fontfamily{phv}\selectfont 
\normalsize \textbf{Data availability} 
}
\newline
\noindent The data that support the findings of this study are available from the corresponding author upon reasonable request.

\vspace{8pt}
\noindent  {\fontfamily{phv}\selectfont 
\normalsize \textbf{Conflict of Interest} 
}
\newline
\noindent The authors declare no competing interests.

\bibliography{references}

@PREAMBLE{
 "\providecommand{\noopsort}[1]{}" 
 # "\providecommand{\singleletter}[1]{#1}%" 
}

@article{kolesov2018superresolution,
  title = {Superresolution {{Microscopy}} of {{Single Rare-Earth Emitters}} in {{YAG}} and $\mathrm{H_3}$ {{Centers}} in {{Diamond}}},
  author = {Kolesov, R. and Lasse, S. and Rothfuchs, C. and Wieck, A. D. and Xia, K. and Kornher, T. and Wrachtrup, J.},
  year = 2018,
  month = jan,
  journal = {Physical Review Letters},
  volume = {120},
  number = {3},
  pages = {033903},
  publisher = {American Physical Society},
  doi = {10.1103/PhysRevLett.120.033903},
}

@article{zhou2024terabitscalea,
  title = {Terabit-Scale High-Fidelity Diamond Data Storage},
  author = {Zhou, Jingyang and Su, Jia and Guan, Junyu and Yang, Yichen and Ji, Wentao and Wang, Mengqi and Shi, Fazhan and Xia, Kangwei and Wang, Ya and Du, Jiangfeng},
  year = 2024,
  month = dec,
  journal = {Nature Photonics},
  volume = {18},
  number = {12},
  pages = {1327--1334},
  publisher = {Nature Publishing Group},
  issn = {1749-4893},
  doi = {10.1038/s41566-024-01573-1},
}

@article{zijlstra2009fivedimensional,
  title = {Five-Dimensional Optical Recording Mediated by Surface Plasmons in Gold Nanorods},
  author = {Zijlstra, Peter and Chon, James W. M. and Gu, Min},
  year = 2009,
  month = may,
  journal = {Nature},
  volume = {459},
  number = {7245},
  pages = {410--413},
  publisher = {Nature Publishing Group},
  issn = {1476-4687},
  doi = {10.1038/nature08053},
}

@article{zhang2014seemingly,
  title = {Seemingly {{Unlimited Lifetime Data Storage}} in {{Nanostructured Glass}}},
  author = {Zhang, Jingyu and Gecevi{\v c}ius, Mindaugas and Beresna, Martynas and Kazansky, Peter G.},
  year = 2014,
  month = jan,
  journal = {Physical Review Letters},
  volume = {112},
  number = {3},
  pages = {033901},
  publisher = {American Physical Society},
  doi = {10.1103/PhysRevLett.112.033901},
}

@article{wang2022100layer,
  title = {100-{{Layer Error-Free 5D Optical Data Storage}} by {{Ultrafast Laser Nanostructuring}} in {{Glass}}},
  author = {Wang, Huijun and Lei, Yuhao and Wang, Lei and Sakakura, Masaaki and Yu, Yanhao and Shayeganrad, Gholamreza and Kazansky, Peter G.},
  year = 2022,
  journal = {Laser \& Photonics Reviews},
  volume = {16},
  number = {4},
  pages = {2100563},
  issn = {1863-8899},
  doi = {10.1002/lpor.202100563},
}

@article{zhao20243d,
  title = {A {{3D}} Nanoscale Optical Disk Memory with Petabit Capacity},
  author = {Zhao, Miao and Wen, Jing and Hu, Qiao and Wei, Xunbin and Zhong, Yu-Wu and Ruan, Hao and Gu, Min},
  year = 2024,
  month = feb,
  journal = {Nature},
  volume = {626},
  number = {8000},
  pages = {772--778},
  publisher = {Nature Publishing Group},
  issn = {1476-4687},
  doi = {10.1038/s41586-023-06980-y},
}

@article{lamon2021nanoscale,
  title = {Nanoscale Optical Writing through Upconversion Resonance Energy Transfer},
  author = {Lamon, S. and Wu, Y. and Zhang, Q. and Liu, X. and Gu, M.},
  year = 2021,
  month = feb,
  journal = {Science Advances},
  volume = {7},
  number = {9},
  pages = {eabe2209},
  publisher = {American Association for the Advancement of Science},
  doi = {10.1126/sciadv.abe2209},
}

@article{gu2014optical,
  title = {Optical Storage Arrays: A Perspective for Future Big Data Storage},
  shorttitle = {Optical Storage Arrays},
  author = {Gu, Min and Li, Xiangping and Cao, Yaoyu},
  year = 2014,
  month = may,
  journal = {Light: Science \& Applications},
  volume = {3},
  number = {5},
  pages = {e177-e177},
  publisher = {Nature Publishing Group},
  issn = {2047-7538},
  doi = {10.1038/lsa.2014.58},
}

@article{kornher2016production,
  title = {Production Yield of Rare-Earth Ions Implanted into an Optical Crystal},
  author = {Kornher, Thomas and Xia, Kangwei and Kolesov, Roman and Kukharchyk, Nadezhda and Reuter, Rolf and Siyushev, Petr and St{\"o}hr, Rainer and Schreck, Matthias and Becker, Hans-Werner and Villa, Bruno and Wieck, Andreas D. and Wrachtrup, J{\"o}rg},
  year = 2016,
  month = feb,
  journal = {Applied Physics Letters},
  volume = {108},
  number = {5},
  pages = {053108},
  issn = {0003-6951},
  doi = {10.1063/1.4941403},
}

@article{kolesov2012optical,
  title = {Optical Detection of a Single Rare-Earth Ion in a Crystal},
  author = {Kolesov, R. and Xia, K. and Reuter, R. and St{\"o}hr, R. and Zappe, A. and Meijer, J. and Hemmer, P. R. and Wrachtrup, J.},
  year = 2012,
  month = aug,
  journal = {Nature Communications},
  volume = {3},
  number = {1},
  pages = {1029},
  publisher = {Nature Publishing Group},
  issn = {2041-1723},
  doi = {10.1038/ncomms2034},
}

@article{kolesov2013mapping,
  title = {Mapping {{Spin Coherence}} of a {{Single Rare-Earth Ion}} in a {{Crystal}} onto a {{Single Photon Polarization State}}},
  author = {Kolesov, Roman and Xia, Kangwei and Reuter, Rolf and Jamali, Mohammad and St{\"o}hr, Rainer and Inal, Tugrul and Siyushev, Petr and Wrachtrup, J{\"o}rg},
  year = 2013,
  month = sep,
  journal = {Physical Review Letters},
  volume = {111},
  number = {12},
  pages = {120502},
  publisher = {American Physical Society},
  doi = {10.1103/PhysRevLett.111.120502},
}

@article{xia2015alloptical,
  title = {All-{{Optical Preparation}} of {{Coherent Dark States}} of a {{Single Rare Earth Ion Spin}} in a {{Crystal}}},
  author = {Xia, Kangwei and Kolesov, Roman and Wang, Ya and Siyushev, Petr and Reuter, Rolf and Kornher, Thomas and Kukharchyk, Nadezhda and Wieck, Andreas D. and Villa, Bruno and Yang, Sen and Wrachtrup, J{\"o}rg},
  year = 2015,
  month = aug,
  journal = {Physical Review Letters},
  volume = {115},
  number = {9},
  pages = {093602},
  publisher = {American Physical Society},
  doi = {10.1103/PhysRevLett.115.093602},
}

@article{sarid2007roadmap,
  title = {A {{Roadmap}} for {{Optical Data Storage Applications}}},
  author = {Sarid, Dror and Schechtman, Barry H.},
  year = 2007,
  month = may,
  journal = {Optics and Photonics News},
  volume = {18},
  number = {5},
  pages = {32--37},
  publisher = {Optica Publishing Group},
  issn = {1541-3721},
  doi = {10.1364/OPN.18.5.000032},
}

@article{herklotz2025modulating,
  title = {Modulating {{Oxide-Based Quantum Materials}} by {{Ion Implantation}}},
  author = {Herklotz, Andreas and Ward, Thomas Zac and Zhou, Shengqiang},
  year = 2025,
  journal = {Advanced Functional Materials},
  volume = {35},
  number = {43},
  pages = {2506647},
  issn = {1616-3028},
  doi = {10.1002/adfm.202506647},
}

@article{ngandeungambou2024hot,
  title = {Hot Ion Implantation to Create Dense {{NV}} Center Ensembles in Diamond},
  author = {Ngandeu Ngambou, Midrel Wilfried and Perrin, Pauline and Balasa, Ionut and Tiranov, Alexey and Brinza, Ovidiu and B{\'e}n{\'e}dic, Fabien and Renaud, Justine and Reveillard, Morgan and Silvent, J{\'e}r{\'e}mie and Goldner, Philippe and Achard, Jocelyn and Tallaire, Alexandre},
  year = 2024,
  month = mar,
  journal = {Applied Physics Letters},
  volume = {124},
  number = {13},
  pages = {134002},
  issn = {0003-6951},
  doi = {10.1063/5.0196719},
}

@article{telkhozhayeva2024roadmap,
  title = {Roadmap toward {{Controlled Ion Beam-Induced Defects}} in {{2D Materials}}},
  author = {Telkhozhayeva, Madina and Girshevitz, Olga},
  year = 2024,
  journal = {Advanced Functional Materials},
  volume = {34},
  number = {45},
  pages = {2404615},
  issn = {1616-3028},
  doi = {10.1002/adfm.202404615},
}

@article{dhomkar2016longterm,
  title = {Long-Term Data Storage in Diamond},
  author = {Dhomkar, Siddharth and Henshaw, Jacob and Jayakumar, Harishankar and Meriles, Carlos A.},
  year = 2016,
  month = oct,
  journal = {Science Advances},
  volume = {2},
  number = {10},
  pages = {e1600911},
  publisher = {American Association for the Advancement of Science},
  doi = {10.1126/sciadv.1600911},
}

@article{monge2024reversible,
  title = {Reversible Optical Data Storage below the Diffraction Limit},
  author = {Monge, Richard and Delord, Tom and Meriles, Carlos A.},
  year = 2024,
  month = feb,
  journal = {Nature Nanotechnology},
  volume = {19},
  number = {2},
  pages = {202--207},
  publisher = {Nature Publishing Group},
  issn = {1748-3395},
  doi = {10.1038/s41565-023-01542-9},
}

@article{gu2016nanomaterials,
  title = {Nanomaterials for Optical Data Storage},
  author = {Gu, Min and Zhang, Qiming and Lamon, Simone},
  year = 2016,
  month = oct,
  journal = {Nature Reviews Materials},
  volume = {1},
  number = {12},
  pages = {16070},
  publisher = {Nature Publishing Group},
  issn = {2058-8437},
  doi = {10.1038/natrevmats.2016.70},
}

@article{zhang20253d,
  title = {{{3D}} Ultra-Broadband Optically Dispersive Microregions in Lithium Niobate},
  author = {Zhang, Bo and Wang, Zhuo and {Albrow-Owen}, Tom and Hasan, Tawfique and Chen, Zesheng and Song, Zhiying and Zhang, Gongyuan and Joyce, Hannah and Tan, Dezhi and Guo, Qiangbing and Qiu, Cheng-wei and Yang, Zongyin and Qiu, Jianrong},
  year = 2025,
  month = jul,
  journal = {Nature Communications},
  volume = {16},
  number = {1},
  pages = {6086},
  publisher = {Nature Publishing Group},
  issn = {2041-1723},
  doi = {10.1038/s41467-025-61317-9},
}

@article{huang2020reversible,
  title = {Reversible 3D Laser Printing of Perovskite Quantum Dots inside a Transparent Medium},
  author = {Huang, Xiongjian and Guo, Qianyi and Yang, Dandan and Xiao, Xiudi and Liu, Xiaofeng and Xia, Zhiguo and Fan, Fengjia and Qiu, Jianrong and Dong, Guoping},
  year = 2020,
  month = feb,
  journal = {Nature Photonics},
  volume = {14},
  number = {2},
  pages = {82--88},
  publisher = {Nature Publishing Group},
  issn = {1749-4893},
  doi = {10.1038/s41566-019-0538-8},
}

@article{sun2022threedimensional,
  title = {Three-Dimensional Direct Lithography of Stable Perovskite Nanocrystals in Glass},
  author = {Sun, Ke and Tan, Dezhi and Fang, Xinyuan and Xia, Xintao and Lin, Dajun and Song, Juan and Lin, Yonghong and Liu, Zhaojun and Gu, Min and Yue, Yuanzheng and Qiu, Jianrong},
  year = 2022,
  month = jan,
  journal = {Science},
  volume = {375},
  number = {6578},
  pages = {307--310},
  publisher = {American Association for the Advancement of Science},
  doi = {10.1126/science.abj2691},
}

@article{lamon2021nanophotonicsenabled,
  title = {Nanophotonics-Enabled Optical Data Storage in the Age of Machine Learning},
  author = {Lamon, Simone and Zhang, Qiming and Gu, Min},
  year = 2021,
  month = nov,
  journal = {APL Photonics},
  volume = {6},
  number = {11},
  pages = {110902},
  issn = {2378-0967},
  doi = {10.1063/5.0065634},
}

@article{neupane2013tuning,
  title = {Tuning Donut Profile for Spatial Resolution in Stimulated Emission Depletion Microscopy},
  author = {Neupane, Bhanu and Chen, Fang and Sun, Wei and Chiu, Daniel T. and Wang, Gufeng},
  year = 2013,
  month = apr,
  journal = {Review of Scientific Instruments},
  volume = {84},
  number = {4},
  pages = {043701},
  issn = {0034-6748},
  doi = {10.1063/1.4799665},
}

@article{ye2025parallel,
  title = {Parallel Writing of {{5D}} Optical Data via Shaped Voxels},
  author = {Ye, Minxin and Lei, Yuhao and Zhang, Xin and Wang, Lei and Chen, Shih-Chi},
  year = 2025,
  month = jul,
  journal = {Science Advances},
  volume = {11},
  number = {29},
  pages = {eadx7335},
  publisher = {American Association for the Advancement of Science},
  doi = {10.1126/sciadv.adx7335},
}

@article{gayen1992twophoton,
  title = {Two-Photon Excitation of the Lowest {$4f^2 \rightarrow 4f5d$} Near-Ultraviolet Transitions in {$\mathrm{Pr^{3+}:Y_3Al_5O_{12}}$}},
  author = {Gayen, S. K. and Xie, Bin-Qing and Cheung, Y. M.},
  journal = {Physical Review B},
  volume = {45},
  number = {1},
  pages = {20--28},
  year = {1992},
  month = jan,
  publisher = {American Physical Society},
  doi = {10.1103/PhysRevB.45.20},
}

@article{jacobs1978measurement,
  title = {Measurement of Excited-state-absorption Loss for {$\mathrm{Ce^{3+}}$} in {$\mathrm{Y_3Al_5O_{12}}$} and Implications for Tunable 5d{$\rightarrow$}4f Rare-earth Lasers},
  author = {Jacobs, Ralph R. and Krupke, William F. and Weber, Marvin J.},
  year = 1978,
  month = sep,
  journal = {Applied Physics Letters},
  volume = {33},
  number = {5},
  pages = {410--412},
  issn = {0003-6951},
  doi = {10.1063/1.90395},
}

@article{allison2026lasera,
  title = {Laser Writing in Glass for Dense, Fast and Efficient Archival Data Storage},
  author = {Allison, James and Anderson, Patrick and Aranas, Erika and Assaf, Youssef and Black, Richard and Caballero, Marco and Canakci, Burcu and Chattaway, John Antony and Chatzieleftheriou, Andromachi and Clegg, James and Cletheroe, Daniel and Cooper, Bridgette and Deegan, Tim and Donnelly, Austin and Drevinskas, Rokas and Feng, Zhonghe and Gkantsidis, Christos and Gomez Diaz, Ariel and Haller, Istvan and Hong, Freddie and Ilieva, Teodora and Joyce, Russell and Kapitany, Valentin and Kunkel, Mint and Lara, David and Lawson, Takashi and Legtchenko, Sergey and Liu, Fanglin and Liu, Xiaoqi and Magalhaes, Bruno and Nowozin, Sebastian and Overweg, Hiske and Rowstron, Antony and Sakakura, Masaaki and Schreiner, Nina and Smith, Adam and Snowdon, Oliver and Stefanovici, Ioan and Sweeney, David and Verkes, Govert and Wainman, Phil and Whittaker, Charles and Wilke Berenguer, Pablo and Williams, Hugh and Winkler, Thomas and Winzeck, Stefan and {Microsoft Research Project Silica Team}},
  year = 2026,
  month = feb,
  journal = {Nature},
  volume = {650},
  number = {8102},
  pages = {606--612},
  publisher = {Nature Publishing Group},
  issn = {1476-4687},
  doi = {10.1038/s41586-025-10042-w},
}

@article{yuan2020ultrahigh,
  title = {Ultra-High Capacity for Three-Dimensional Optical Data Storage inside Transparent Fluorescent Tape},
  author = {Yuan, Xupeng and Zhao, Miao and Guo, Xinjun and Li, Yao and Yu, Yang and Gan, Zongsong and Ruan, Hao},
  year = 2020,
  month = mar,
  journal = {Optics Letters},
  volume = {45},
  number = {6},
  pages = {1535--1538},
  publisher = {Optica Publishing Group},
  issn = {1539-4794},
  doi = {10.1364/OL.387278},
}

@inproceedings {315322,
author = {Qing Wang and Fan Yang and Qiang Liu and Geng Xiao and Yongpeng Chen and Hao Lan and Leiming Chen and Bangzhu Chen and Chenrui Liu and Pingchang Bai and Bin Huang and Zigan Luo and Mingyu Xie and Yu Wang and Youyou Lu and Huatao Wu and Jiwu Shu},
title = {Cost-efficient Archive Cloud Storage with Tape: Design and Deployment},
booktitle = {24th USENIX Conference on File and Storage Technologies (FAST 26)},
year = {2026},
isbn = {978-1-939133-53-3},
address = {Santa Clara, CA},
pages = {19--34},
url = {https://www.usenix.org/conference/fast26/presentation/wang},
publisher = {USENIX Association},
month = feb
}
 
% \vspace{8pt}
% \noindent  {\fontfamily{phv}\selectfont 
% \normalsize \textbf{Acknowledgements} 
% }
% \newline
% \noindent This work is supported by the National Natural Science Foundation of China (12274400, 12504594, T2325023, 92265204, 123B1019), the Strategic Priority Research Program of the Chinese Academy of Sciences (XDA0590204, XDB1740301), Quantum Science and Technology-National Science and Technology Major Project (2021ZD0302200), and the Fundamental Research Funds for the Central Universities (WK2030000076). The nanofabrication was carried out at the USTC center for Micro and nanoscale Research and fabrication.

% \vspace{8pt} 
% \noindent  {\fontfamily{phv}\selectfont 
% \normalsize \textbf{Author contributions} 
% }
% \newline
% \noindent K.X., J.G., and Q.S. designed the experiment. Q.S., J.G., and H.Z. prepared the sample. J.G., Q.S., H.W., Z.C., Z.G. and K.X. performed the data-reading experiments and data analysis. B.T. and D.L. performed the algorithmic analysis. J.G., K.X., D.L., B.T., Q.S., Y.W., Z.C., J.Z., and Z.G. wrote the paper.  K.X. supervised the project and conceived the idea. All authors discussed the results and commented on the paper.

% \vspace{8pt}
% \noindent  {\fontfamily{phv}\selectfont 
% \normalsize \textbf{Conflict of Interest} 
% }
% \newline
% \noindent The authors declare no competing interests.

\end{document}